# Distributed Accountability in Democracy: Using MANETs and DTNs in the Face of Acts of Questionable Legality

Mathew Schmidheiser
*School of Computer Science*
*The University of Nottingham*
psxms34@nottingham.ac.uk

Milena Radenkovic
*School of Computer Science*
*The University of Nottingham*
milena.radenkovic@nottingham.ac.uk

*Abstract*— In this paper, we explore the behavior of the Epidemic and Wave DTN routing protocols in a realistic setting where individuals may wish to communicate with others for support regarding an act of questionable legality. We identify situations where using the Epidemic routing protocol may be more advantageous in such a scenario, and situations where using the Wave routing protocol may be more advantageous instead. We discuss other aspects of our findings in detail and suggest multiple approaches to future works.

*Keywords* — *DTN, MANET, Distributed Accountability, Human rights, Essential communication in crowded environments*

## I. INTRODUCTION

Accountability is a fundamental component of democracy. When questions arise regarding the legality, violation of rights, or other sensitive characteristics of actions taken by the state, the state's populace may wish to voice their opposition to the action or record information about action instances.

Efficient local, mobile communication is critical to this ability. When witnessing an event of questionable legality or a potential violation of rights, an individual may wish to quickly request support recording information from others nearby; when voicing their opposition through protest, individuals may need to rapidly organize or communicate critical safety information to a vast number of attendees. Congestion in cellular networks can render traditional mobile communication unavailable [1], or the networks may be disabled by the state [2], [3].

This work explores the use of delay tolerant network (DTN) routing protocols in a dynamic bluetooth mobile ad hoc network (MANET) of mobile consumer edge devices to efficiently disperse a message throughout the network. To do this, we use the Opportunistic Network Environment (ONE) simulator to analyze the performance of the Epidemic [4] and Wave [5] DTN routing protocols in networks with varying numbers of nodes and message creation frequencies.

We present the remainder of this work as follows:

- In Section 2 we discuss the background and motivation for this work. This includes a critical discussion and analysis of MANETs, DTNs, the Epidemic routing protocol, and the Wave routing protocol; a discussion of the ONE simulator; a discussion of existing solutions to our presented problem; and a review of current works related to our experiment;

- In Section 3 we present our simulation design and architecture;

- In Section 4 and Section 5 we present our results and discussion, respectively;

- In Section 6 we present further recommendations for future research;

- In Section 7 we discuss works exploring other uses of MANETs and DTNs; and

- In Section 8 we present our conclusion.

## II. BACKGROUND AND MOTIVATION

In this work, we aim to explore how a message may be most efficiently delivered at least one time to every node in a MANET using a DTN protocol. Through this exploration, we aim to identify the degree to which a routing protocol meets the following criteria:

- The protocol can distribute a message to every node in a network. We refer to this as "saturating" a network with a message, and when a message has been delivered to each node in the network at least once we say the network has been "saturated". Note that we are not using "saturated" to refer to buffer occupancy volume or network congestion, but rather solely to refer to the number of unique nodes to which a message has been delivered.

- The protocol can saturate a network with a message in a short amount of time.

- The protocol can saturate a network with a message with low network and device resource expenditure.

We explore the performance of the Epidemic routing protocol [4] and Wave routing protocol [5] to evaluate them against these criteria. We discuss the details of these protocols and why we selected them, as well as of MANETs, DTNs, and further relevant information, throughout this section.

### A. Mobile Ad Hoc Networks (MANETs)

A mobile ad hoc network (MANET) is a distributed network in which potentially heterogeneous participants, referred to as nodes, are typically physically moving, act as both hosts and routers, and communicate amongst each other without centralised infrastructure [6], [7]. Nodes engage in wireless peer-to-peer communication, and each helps route data to the correct destination [6], [7]. Semi-mobile or stationary nodes may also be present [7]. Put simply, a MANET is a decentralised, frequently-changing network in which network nodes, which are typically physically mobile, both receive messages from and help send messages to other nodes in the network to omit the need for centralised communication infrastructure.

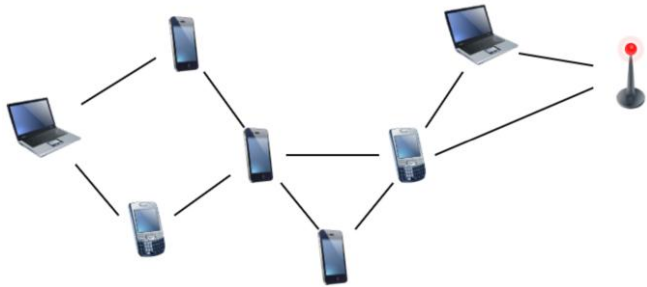

Fig. 1. Example MANET with heterogeneous nodes. MANETs may optionally include fixed nodes as well, as shown in this diagram. As MANETs have dynamic network topologies, this diagram represents just a snapshot of an example MANET at a given time.

The mobility of MANET nodes results in a highly-dynamic network topology, and accordingly challenges can arise with message routing and network congestion [6], [7]. Routing challenges can also be posed by nodes' limited compute, capabilities, and battery capacity compared to the infrastructure of traditional internet networks [6], [7].

Three main types of ad hoc routing protocols are used in MANETs: on-demand routing protocols, table-driven routing protocols, and hybrid routing protocols [7].

*1) On-Demand Routing Protocols*

Also called reactive protocols or source-initiated protocols, the primary feature of on-demand routing protocols is that nodes only identify routes to a destination when necessary [6], [7]. In other words, nodes do not maintain realtime knowledge of the network topology; instead, they evaluate the network topology only when they need to send a message, and during evaluation only seek to identify topology relevant to their message's destination.

How the best route to a destination is discovered and determined depends upon the specific on-demand routing protocol used. For example, the Associativity-Based Routing (ABR) protocol determines the best route to a destination by identifying the route which contains nodes that have been in longest continuous contact with each other (i.e. the most "stable" route) [6], [8], [9].

On-demand routing protocols help save network bandwidth by having nodes only send route identification-related traffic when a message must be sent, although congestion from highly-frequent performance of route discovery can occur in extreme situations [7]. Additionally, performing route identification at the time of message transmission can increase message delivery time.

*2) Table-Driven Routing Protocols.*

Also called proactive routing protocols, networks using table-driven routing protocols require nodes maintain real-time knowledge of the network topology [6], [7]. Knowledge is maintained in the form of periodically-exchanged routing tables, and a node will reference its routing table to identify the best route to deliver a message to a destination when needed [6], [7].

How routing tables are exchanged and how the best route for a message is determined once again depends upon the specific routing protocol used. For example, nodes in a network using the Destination-Sequenced Distance-Vector (DSDV) protocol continuously exchange periodic routing table updates, and when a node identifies significant new information concerning its routing table it shares an update immediately [6], [10].

While networks which implement a table-driven routing protocol do not need to spend time performing route identification at message transmission time, as routes are already identified in nodes' routing tables, the effort required to keep routing tables updated increases network traffic, potentially significantly in highly-dynamic networks [6], [7]. Accordingly, these types of protocols are best suited for networks which change infrequently [6], [7].

*3) Hybrid Routing Protocols.*

Protocols of this type combine aspects of both table-driven routing protocols and on-demand routing protocols to leverage the advantages of each [6], [7]. Which aspects of the protocol types are implemented and how the implementations are performed depends upon the specific hybrid routing protocol. For example, the Zone Routing Protocol (ZRP) uses routing tables for communication with local nodes and performs route discovery when a message must be sent outside of a node's local area [6], [11].

A wide variety of MANET applications exist, such as emergency services networks, mobile conferencing, embedded computing applications, and sensor networks [6]. Additionally, while outside the scope of this work, Vehicular Ad Hoc Networks (VANETs), a special type of MANET, can be used to communicate critical safety information, traffic information, and entertainment in networks oriented around transportation vehicles [6].

Appropriate MANET routing protocol selection can help mitigate the challenges encountered in MANETs discussed at the beginning of this section. However, in situations where message delivery must be able to overcome disconnections or "gaps" in the network, a delay tolerant network (DTN) should be implemented [6], [12].

*B. Delay Tolerant Networks (DTNs)*

Delay tolerant networks (DTNs) enable communication in situations where an end-to-end connection may not exist between a source node and a destination [6], [12]. Rather than promptly forwarding a message upon receipt as in traditional internet networks and MANETs, a node in a DTN can store a message for longer periods of time and forward it once an appropriate opportunity arises in what is called the "store-carry-forward" paradigm [6], [12]. This paradigm can be used to enable networking in a variety of challenging and sparse communication environments, including interplanetary networks, military networks, intermittent sensor networks, disaster response networks, and other environments in which network topology may be sparse or communication may otherwise be difficult to perform immediately [6], [12], [13], [14].

Routes must be identified with care in DTNs, as even just one poor forwarding decision can greatly exacerbate latency or result in a message failing to be delivered entirely [6]. There are two main types of DTN routing protocols: forwarding-based protocols and replication-based protocols [12], [15].

*1) Forwarding-Based Protocols*

Forwarding-based protocols involve one copy of a message being forwarded from the source node to destination node, potentially through a series of intermediary nodes [12], [15].

How the node to which a message should be forwarded is determined depends on the specific forwarding-based protocol. For example, in the First Contact protocol, a node forwards a message to the first node it encounters, while in the Direct Delivery protocol a node forwards a message only directly to the destination node [16].

Forwarding-based protocols generally use less network resources, but often yield inadequate message delivery rates due to the challenge of successfully routing a message through the network using only one copy [12], [17], [18], [19].

*2) Replication-Based Protocols*

Replication-based protocols involve nodes distributing multiple copies of a message throughout the network, resulting in a higher message delivery probability [12]. Rather than just forwarding a message to one node as occurs in forwarding-based protocols, a node can forward multiple copies of a message as it encounters other nodes.

How many times a message is replicated and how replicas are routed through the network varies between replication-based protocol implementations. For example, in the Epidemic protocol, which we describe further in the next section and use in this work, a node often attempts to deliver a replica to every other node it comes into contact with, while the Spray and Wait protocol limits the number of replicas of a message which may be made and will only deliver a message via Direct Delivery once the limit is reached [4], [12], [18].

While replication-based protocols generally have higher message delivery rates than forwarding-based protocols, they can also consume significant network resources, especially when naïve distribution is performed, and some implementations are inherently unscalable [12], [15].

### C. Epidemic Routing Protocol

The Epidemic routing protocol is a DTN protocol that attempts to deliver a message in a disconnected network through widespread message replication [12]. When a node has a message to deliver, it sends a replica of the message to every other node it comes into contact with, with the idea that as nodes physically move throughout the network and continue to spread the message in the same fashion the message will eventually be delivered to its destination [4], [6], [12].

To reduce the use of node and network resources, nodes keep a cache of recentlycontacted nodes and only initiate message exchanges with other nodes that they have not recently come into contact with [4], [6], [12]. Additionally, when message exchange does occur, nodes first exchange vectors indicating the messages they currently possess [4], [6]; then, each node requests only unpossessed messages from the other [4], [6].

Each node stores the messages it possesses in its message buffer [4]. When a node's message buffer reaches capacity and it is to receive a new message, it removes existing content from the message buffer to make room for the new message [4]. Various types of algorithms can be used to remove existing content from the message buffer [4]; for example, when using a First In First Out (FIFO)-style algorithm, the oldest content in the buffer is removed to make room for new content [4]. After a message is removed from a node's buffer, the node is eligible to receive that message again.

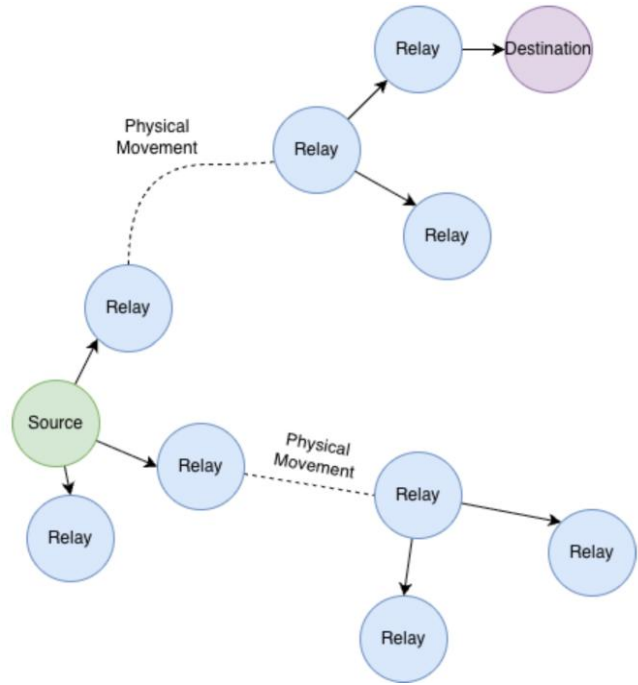

Fig. 2. Diagram of the Epidemic protocol distributing a message throughout a network, with the message eventually reaching its destination.

While the Epidemic protocol enjoys a high message delivery rate [4], its naïve distribution of messages can use large amounts of resources and significantly degrade network performance [15], [18], [19]. Specifying a maximum number of times a message may be replicated to a node other than its destination before it will only be delivered via Direct Delivery, as the Spray and Wait protocol does for example, can help alleviate this issue [4], [18].

The traditional Epidemic protocol was designed for the purpose of delivering a message to a specific destination [4]. However, as messages are able to be widely dispersed throughout the network as a part of the delivery process, we can also use the protocol to attempt to saturate a message throughout a network and measure the results.

The Epidemic protocol can also be modified for the specific purpose of relaying a message throughout a network, i.e. "broadcasting" the message [2], [4], [20]; we leave implementing this variation of Epidemic in the ONE simulator for a future work.

### D. Wave Routing Protocol

The Wave routing protocol is an Epidemic-like protocol that aims to circulate messages in a network [5], [21]. Like in an Epidemic network, nodes in a network implementing the Wave protocol receive messages from other nodes, replicate messages they possess to other nodes, and delete messages they possess after a set criteria is met [5]. However, Wave differs from Epidemic in a significant way important to this work: in a Wave network, a node may refuse to accept a message it has already received even after the message has been deleted from its buffer [5].

To implement this refusal behavior, each node maintains a list in which it tracks messages it has received within a specified previous period of time [5]. The list is called a tracking list, and the specified period of time is called the immunity time [5]. When a node exchanges messages with another, it will not request any message which is in its tracking list, even if the message is not present in the node's buffer [5].

A message is removed from a node's buffer in two situations. The first is when the message's time-to-live (TTL), a message-specific value which specifies the amount of time for which a message should be distributed after its creation, expires [5]. The second is when the custody time, a node-specific value which specifies the maximum amount of time a node will keep a message in its buffer, expires [5]. The custody time is a specified fraction of the node's immunity time [5].

Unlike traditional Epidemic, the goal of Wave routing involves delivering a message throughout a network. Specifically, Wave aims to deliver a message in "waves" throughout a grid [5]. Accordingly, it is appropriate to evaluate it against our three criteria.

*E. The ONE Simulator*

The Opportunistic Network Environment (ONE) simulator is Java software originally developed at Aalto University designed to simulate the performance and behavior of DTN routing protocols [16], [22]. The simulator allows a user to design a network topology using data that represents physical pathways, such as footpaths or roadways; data describing a movement model, such as random movement; and data describing the characteristics of network nodes, such as their quantity. A user may then use DTN routing protocol emulators programmed into the software to simulate protocol behavior, and data outputs generated during simulation may be analyzed to evaluate performance. A detailed description of this functionality and further functionality discussed below is available in [23].

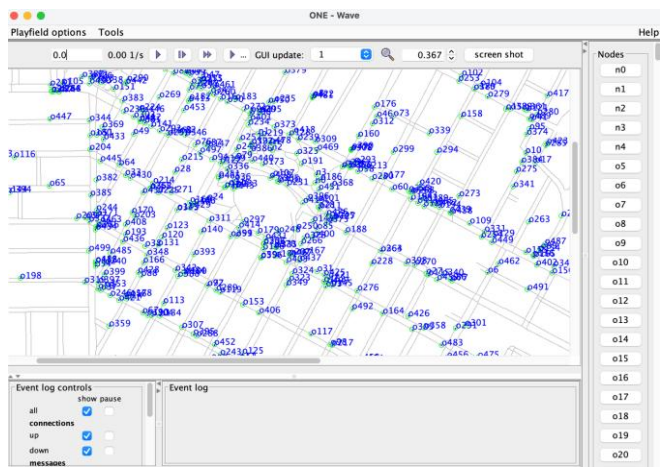

Fig. 3. The ONE simulator run in GUI mode. Map data © OpenStreetMap contributors.

The simulator may be run with a graphical user interface (GUI) to execute a single simulation and visually see node behavior. It may also be run from the command line to automatically execute multiple simulations with varying simulation parameters in sequence; this is called "batch mode" [24].

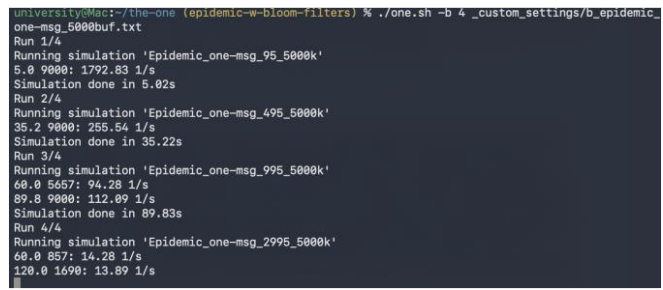

Fig. 4. The ONE simulator run in batch mode.

Simulation parameters, such as routing protocol implementation, number of nodes, and mobility models, are specified in ONE configuration files. ONE configuration files arespecially formatted .txt files read by ONE at simulation time to construct the simulation. Many of the simulation parameters are described in the ONE simulator's README.md file that is a part of the software's source code and is also viewable on the software's GitHub repository [24]. All of the simulation parameters are present in their respective locations of the software's source code files; for example, parameters relevant to the simulator's implementation of the Epidemic routing protocol may be found in the EpidemicRouter.java file. The specific simulation parameters we used in this work are described in Section 3.

A notable limitation of the ONE simulator is that it does not fully model all lower-layer network behavior [16]. For example, ONE does not model signal attenuation nor physical medium congestion [16]. Inyangson and Radway et al's work identifies that simulating lower-layer network characteristics and behavior can show degraded protocol performance compared to simulating only higher-layer network characteristics and behavior [2]. Accordingly, we note that further simulations which more accurately model lower-layer network characteristics and behavior should be performed to produce a further realistic understanding of DTN protocol behavior in the scenario explored in this work. We leave this for future research.

*F. Existing Solutions*

A variety of technologies which may be used in the scenario we explore in this work exist presently or have been used previously. We now discuss a selection of these technologies and their strengths and limitations in detail.

*1) Cellular network-based solutions.*

Technologies which leverage traditional cellular networks, such as the Signal mobile application [25], can be used in the scenario we explore in this work. Signal in particular is a communication app which implements end-to-end encryption to protect user privacy [26]. It is also already recommended for use in the situation of protests against the state [27], [28], [29]. However, cellular network-based solutions are only able to work effectively if the cellular network they require works effectively, and this is not always the case. Cellular network performance can become severely degraded in densely populated environments [1] or be disrupted by state-based actors [30]. Additionally, as discussed by Inyangson and Radway et al, state-based actors may target the internet technologies used by the communication technologies [2], [31] or shut down internet access entirely [2], [3]. Due to its decentralised and delay-tolerant nature, a MANET which implements a DTN routing protocol may be more resilient to these types of challenges. However, it is known that such a network is still vulnerable to network congestion [2].

*2) Previously discussed mesh network-based solutions*

Related works identify existing mesh network-based solutions, such as Bridgefy [2], [20], [32]. Presently, Bridgefy's website advertises that the app can be used at protests [32]. However, related works discuss that the app has previously failed to handle large amounts of protest traffic [2] and has also failed to protect users users' anonymity [2], [20]. These identifications make it seem an inadequate solution for the scenario we are exploring in this work. We leave the exploration of other mesh network-based solutions discussed in related works, such as FireChat [20], for future works, and instead explore a new mesh-network based solution, BitChat, in greater depth.

*3) BitChat*

BitChat is a mobile application released in July 2025 that establishes a Bluetooth mesh network and uses it to transmit encrypted communications between nodes [33], [34].

The app implements the "BitChat Protocol," [34] which is a type of combined protocol that uses either the Direct Delivery protocol or an Epidemic-like protocol depending on the situation. Using the app, users can send a "private message" [34] with a destination of a specific node or a "broadcast message" [34] which is to be distributed to all nodes in the network [34]. When the destination node is within communication range, nodes exchange messages via the Direct Delivery [34]. When the destination node is not within communication range, messages are routed using an Epidemic-like protocol which BitChat titles "Efficient Gossip with Bloom Filters" [34]. In this approach, the message is widely distributed, or "flooded" [34], throughout the network [34]. However, rather than exchanging summary vectors to identify messages a node possesses like traditional Epidemic, a node in BitChat's network stores the IDs of packets it receives in a nodespecific Bloom Filter [34]. The node then evaluates the filter when it receives a new packet to identify if the packet is a duplicate [34].

While BitChat was released only recently, it has already been used by protestors at largescale protests against the Nepali government [35], [36]. Considering it uses a distributed Bluetooth mesh network, implements an Epidemic-like protocol used to widely distribute messages, and was designed for secure communication [34], we have integrated characteristics of the BitChat Protocol into the networks we simulate in this work to evaluate their efficacy and further emulate a relevant network which may be found in the real-world. The integrated characteristics are identified in detail in Section 3.

*G. Related Works*

In their work Strong Anonymity for Mesh Messaging [20], Perry et al discuss a secure, Epidemic-like messaging protocol which aims to distribute a message throughout a network and uses bitstring-based digests, rather than traditional summary vectors, to facilitate message exchange between nodes. When nodes begin the message exchange process, they first exchange digests [20]; then, if a node identifies that the other possesses messages it does not have, it requests the messages [20]. Perry et al also propose an extension to this protocol that organizes nodes into cliques [20]. Each clique has a leader, and leaders facilitate broadcasting messages from their clique members to other cliques and distributing messages received from other cliques to their clique members [20].

In their work Amigo: Secure Group Mesh Messaging in Realistic Protest Settings, Inyangson and Radway et al evaluate the performance of multiple existing DTN routing protocols and one novel DTN routing protocol in various simulated protest scenarios using the ns-3 simulator [2], [37], [38]. Specifically, they evaluate a variation of Epidemic designed to distribute a message throughout a network, an Epidemic-like protocol which uses Bloom Filters to perform the function of summary vectors as proposed by Perry et al [20], a "static clique routing" [2] protocol based on the concept of clique graphs proposed by Perry et al [20], and a novel "dynamic clique routing" [2] protocol which augments Perry et al's clique routing by allowing cliques and clique leaders to change over time [2].

Unlike the ONE simulator, the ns-3 simulator is capable of modelling lower-level network layer behavior, including Layer 1 [2], [37]. In addition to identifying high performance of the dynamic clique routing protocol, Inyangson and Radway et al also identify that lower-level network behavior, such as Layer 1 behavior, plays a crucial role in network performance, and that further research should be conducted in this area. Accordingly, we again note that further research into the scenario we explore should be conducted in a way that explores lower-level network behavior and performance unable to be captured by the ONE simulator.

There are also many works which discuss the use of DTNs in disaster scenarios, such as Obayashi and Ohta's work proposing a protocol with a broadcast-like goal of distributing information in a network during an evacuation [39]. We leave review of such further related works for future research.

III. SIMULATION DESIGN

For our simulation, we consider the following hypothesis: When messages begin to be removed from node buffers, it will take longer for messages in an Epidemic network to be delivered to each unique node at least once than in a Wave network where the immunity time is greater than message TTL, as nodes in the Epidemic network may receive removed messages multiple times inducing congestion, but nodes in the Wave network will receive each message only once. Accordingly, we conduct simulations with varying node quantities, node buffer sizes, and message creation frequencies for both network types. We now describe the technical characteristics of our parameter implementations and our practical simulation scenario in detail.

*A. Routing Protocols*

*1) Epidemic*

The Epidemic routing protocol was implemented with the ONE simulator's EpidemicRouter class. The EpidemicRouter class implements the traditional Epidemic routing protocol with a drop-oldest approach to removing messages from a buffer when it is at capacity and the limitation that a node may only engage in one message transfer at a time [40].It also allows for a time-based message time-to-live (TTL). We use a TTL of 60 minutes for both protocols and discuss this value later in this section.

*2) Wave*

The Wave routing protocol was implemented with the ONE simulator's WaveRouter class. The WaveRouter implements the Wave routing protocol in alignment with [5], discussed in Section 2. To evaluate the effect of immunity time in situations where message removal from a node's buffer may occur, we use an immunity time equal to our simulation time to avoid message re-deliveries. We also use a custody time fraction of 0.5, which makes the custody time longer than message TTL, so that messages are only removed from a nodes buffer due to either the buffer reaching maximum capacity or a message TTL expiring. We recommend future works explore the impact of a custody time which is less than message TTL on the results of this experiment.

### B. Node and Message Characteristics

Node characteristics integrate characteristics of the BitChat Protocol to reflect current technological capabilities and characteristics leveraged in previous works. Characteristics are now described in detail.

*1) Wireless Interface*

A Bluetooth-like interface is simulated as the wireless interface, as the BitChat Protocol Whitepaper specifically discusses the use of Bluetooth Low Energy (BLE) [34]. A transmission radius of 10 meters and a transmission speed of 1.4MB/s is used in alignment with Perry et al.'s representation of a Bluetooth-like interface [20].

*2) Buffer Size*

In alignment with previous works [2] and to explore the impact of various message creation frequencies on buffer capacity, we explore node buffer capacities of 500KB and 5000KB. We expect to see more drastic differences between the behavior of the two routing protocols in the 500KB buffer simulations due to more messages being removed from buffers.

*3) Message Size*

Messages are 2064 bytes in size to simulate a potential maximum size of a message sent using the BitChat Protocol. The BitChat Protocol exchanges data between nodes by sending a binary packet as the payload of a Noise transport message [34]. The binary packet is padded up to a size of 2048 bytes to obfuscate its true size, and a Noise transport message contains 16 bytes of authentication data in addition to its payload [34], [41].

The BitChat Protocol also includes a fragmentation protocol to manage transmission limits of transport layer protocols [34]. Effects of this protocol and transport layer protocol limitations more broadly on practical message size, and consequently message saturation, is not considered in the scope of this work and is left for future study.

*4) Message Time-To-Live (TTL)*

After considering multiple values, we decided to explore a TTL of 60 minutes, as in practical applications it may be important for new nodes entering a network to receive a message a notable amount of time after the message was initially created. For example, if pepper spray or a chemical agent such as tear gas is dispersed at a location, individuals who approach the location later may wish to be aware of the event to avoid any remaining chemical agent.

### C. Scenario: Varying Node Quantities and Message Creation Frequencies at a Los Angeles Park

*1) Location*

United States Immigration and Customs Enforcement (ICE) has been scrutinized for actions deemed to be of questionable legality and/or a violation of rights in cities around the United States [42], [43]. As a part of its operations, the agency has conducted enforcement operations in public locations, including at MacArthur Park in Los Angeles, California [44]. While no actions of questionable legality to individuals were found to be reported occurring as a part of this operation, the location provides a setting for analysis if an individual may have wanted to ask others nearby for help performing legal methods of recording the event.

*2) Location Emulation*

OpenStreetMap (OSM) data of MacArthur Park and a small surrounding area was acquired using the QuickOSM plugin of the QGIS geographic information system software [45], [46], [47]. QGIS was used to export the data in WKT format, and a custom Python scripted was used to combine the WKT data into files for use in The ONE simulator with support from [48], [49]. An OSM basemap was acquired using the QuickMapServices QGIS plugin [50] to serve as a point of reference throughout the process. The custom Python script used to combine WKT data will be available on our GitHub repository.

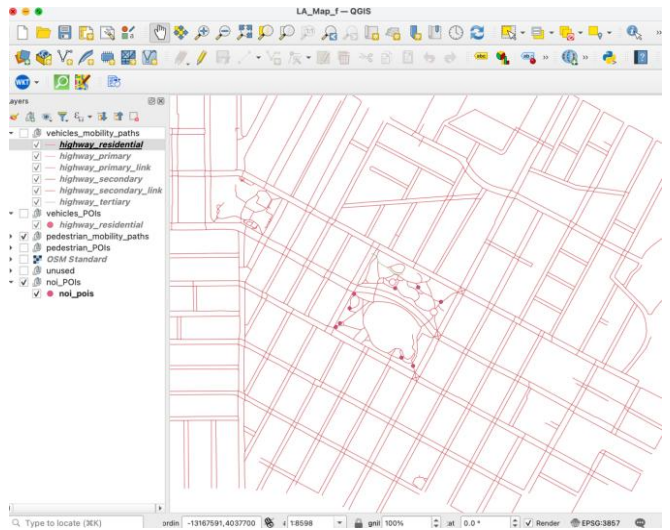

Fig. 5. MacArthur Park map data in the QGIS geographic information system software. Map data © OpenStreetMap contributors.

*3) Node Groups*

Two groups of nodes were used to simulate this scenario.

Group 1 includes nodes we call Nodes of Interest (NOIs). These nodes represent nodes that other, non-NOI nodes may be interested in. In this scenario specifically, they represent nodes conducting an action of questionable legality. Five nodes exist in this group, and each one represents a collection of NOI actors. For example, if 20 actors are a part of an action of questionable legality, they may split into five groups of four actors each; thus, this is represented by five nodes. This principle can be applied to represent any number of NOI actors as five nodes in this group.

Group 2 includes nodes all other nodes which are not NOIs (non-NOIs). In this scenario specifically, they represent bystanders to the action of questionable legality and other individuals in the local area who may want to exercise legal means to document the event. Simulations were conducted with four different quantities of nodes in Group 2 to explore how node quantity impacts time to complete message saturation. These quantities were 95 nodes, 495 nodes, 995 nodes, and 2995 nodes to explore low extremes, high extremes, and intermediary values.

Additionally these perhaps odd-seeming quantities of Group 2 nodes were selected for evaluation purposes to create a desired total quantity of nodes when the number of nodes in Group 1 and Group 2 are combined. For example, 5 NOI nodes and 95 non-NOI nodes sum to 100 total nodes. This is discussed further in the Message Creation subsection of this section.

Nodes representing vehicles or other non-pedestrian entities are not included in this scenario. Including such nodes and their impact on message saturation is an opportunity for further research.

*4) Mobility Patterns*

Both Group 1 nodes (NOIs) and Group 2 nodes (non-NOIs) use The ONE simulator's ShortestPathMapBasedMovement model. In this model, nodes are constrained to a specified map area and travel to random locations within the area using the shortest path as identified by Dijkstra's algorithm [24].

In our scenario, once a node reaches its destination, it waits for a random time between 0 and 120 seconds in alignment with the default ONE configuration settings. It then begins traveling to a new location. Nodes move at varying speeds between 1.31m/s to 1.72m/s based on a United States National Institute of Health (NIH) study which identified these values as the "usual" and "fast" level walking paces for apparently healthy adults [51].

Modifying these settings to more closely reflect specific mobility patterns which may be present during an action of questionable legality and identifying the impact on message saturation is left as an opportunity for further research.

The map area in which Group 1 nodes (NOIs) may travel is confined to MacArthur Park and excludes the surrounding area to simulate NOIs conducting activity in the park. Group 2 nodes (non-NOIs) may travel the entire map area, i.e. both MacArthur park and the surrounding area.

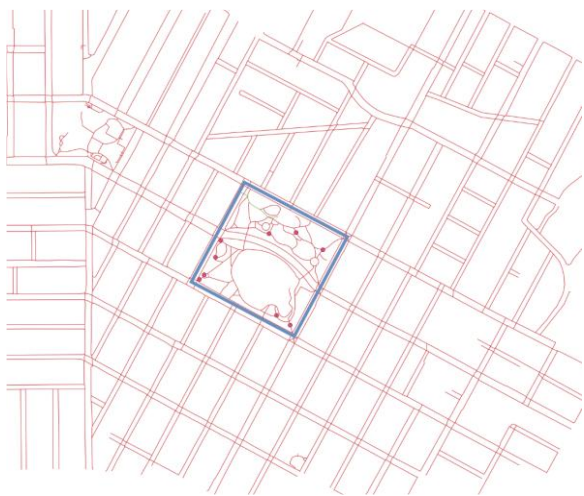

Fig. 6. MacArthur Park map data. Group 1 (NOI) nodes may only travel within the park, indicated by the blue square. Group 2 (non-NOI) nodes may travel anywhere on the map. Map data © OpenStreetMap contributors.

*5) Message Creation*

In the context of message creation, distribution, and saturation, each Group 1 node (NOI) represents a non-NOI actor that is in extremely close proximity to the NOI and creates a message. For example, a bystander may approach an NOI actor, identify an act of questionable legality occurring, and create a message asking for help performing legal means of documenting the event. With this concept in mind, messages are created by Group 1 nodes in the simulation and may have a destination of a Group 1 node or a Group 2 node.

To evaluate the impact of message creation frequency on the time it takes for a message to saturate all nodes, messages are created at three different rates:

- One Message Frequency: 1 message is created at the beginning of the simulation and no further messages are created afterwards. This serves as a baseline measure and seeks to answer the question "Is it possible to efficiently distribute a message amongst this quantity of nodes at all?". We also refer to these simulations as One Message Simulations.

- Moderate Message Frequency: 1 message per NOI is created at the beginning of the simulation; then, 1 message is created per NOI at least every 5 minutes. This serves to simulate somewhat frequent messages being created about the NOIs; for example, over the course of an instance of NOI activity, a bystander may create a new message asking for documentation support when an NOI begins engaging in new behavior. We also refer to these simulations as Moderate Message Simulations.

- High Message Frequency: 1 message per NOI is created at the beginning of the simulation; then, 1 message is created per NOI at least every 30 seconds. This serves to simulate frequent messages being created about NOIs by potentially multiple bystanders; for example, a collective of bystanders may all frequently create messages asking for documentation support. This message creation rate aims to test the limits of conditions in which efficient message saturation is possible. We also refer to these simulations as High Message Simulations.

*6) The ONE Configuration Settings*

We will release our ONE configuration files after publication. We recognize the great help of the following resources in configuring ONE simulations: [22], [24], [52], [53], [54], [55].

*D. Evaluation Method*

Message delivery efficiency is measured by analyzing the time it takes for a message to be delivered to all nodes. While a message has one specific node identified as its destination, the Epidemic and Epidemic-based routing protocols explored in this work involve all nodes distributing a message to all other nodes they come into contact with. Thus, whether or not a node is a message's destination, the node has the potential to receive a copy of the message, and therefore the time it takes a message to be delivered to all nodes can be measured.

IV. RESULTS

The ONE Simulator's Event Log Report was used to record simulation data. The data was analyzed using custom Python scripts to evaluate message saturation rate and the time to total message saturation. The Python scripts will be available on our GitHub repository.

We configured a simulation of 9000 seconds (2.5 hours) and allowed messages to be created for the first 60 minutes. This allowed for 60 minutes of message creation, an additional 60 minutes of message delivery before all message TTLs expired, and 30 minutes of buffer and cooldown time. We considered multiple simulation runtimes and believed this to be an appropriate balance that will allow us to produce and evaluate situations in which behaviors between the Epidemic and Wave networks will differ due to node buffer occupancy while also not running excessively long, high-compute simulations.

We now present our results in detail in three categories divided by message creation frequency.

*A. One Message Frequency Simulations*

1 message at the start of the simulation. Figure 7 shows the message saturation percentage (number of unique deliveries / total nodes in the simulation) by elapsed simulation time for 100, 500, 1000, and 3000 nodes with buffer sizes of 500KB using the Epidemic routing protocol. Figure 8 shows the same but using the Wave routing protocol.

Epidemic (500KB Buffers)

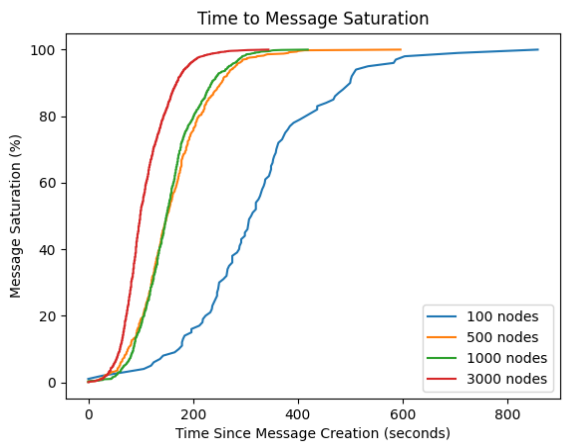

Fig. 7. Message saturation percentage by elapsed simulation time since message creation for with 1 message created at the beginning of the simulation with the Epidemic protocol.

Wave (500KB Buffers)

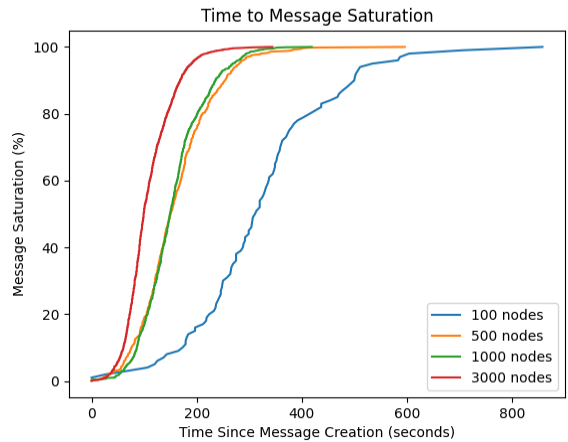

Fig. 8. Message saturation percentage by elapsed simulation time since message creation for with 1 message created at the beginning of the simulation with the Wave protocol.

We can see that the graphs share the same shape. In all simulations, the single transmitted message eventually reached 100% saturation, i.e. it was eventually delivered to all nodes in the network. The same amount of time elapsed between message creation and 100% saturation for each corresponding simulation between the two protocols, as can be seen in Table 1.

TABLE I. ELAPSED SIMULATION TIME FROM MESSAGE CREATION TO 100% MESSAGE SATURATION FOR WITH 1 MESSAGE CREATED AT THE BEGINNING OF THE SIMULATION. TIMES ARE ROUNDED TO ONE DECIMAL PLACE.

|  |  | 100 Nodes | 500 Nodes | 1000 Nodes | 3000 Nodes |
|---|---|---|---|---|---|
| Time to 100% Message Saturation (s) | Epidemic | 858.3 | 595.9 | 418.9 | 344.0 |
|  | Wave | 858.3 | 595.9 | 418.9 | 344.0 |

As we expect to see differences between the performance of our implementations of the Epidemic and Wave protocols when messages begin to be removed from node buffers, it is unsurprising that this identical behavior occurs when only one 2064 byte message is transmitted during the simulations. Such a situation will result in buffers not even beginning to approach their capacity, as is shown in Table 2.

TABLE II. MAX AVERAGE BUFFER CAPACITY BY NODE COUNT AND ROUTING PROTOCOL WITH NODES WITH 500KB BUFFERS IN SCENARIO A.

|  |  | 100 Nodes | 500 Nodes | 1000 Nodes | 3000 Nodes |
|---|---|---|---|---|---|
| Max Average Buffer Occupancy (% of 500KB) | Epidemic | 0.4087 | 0.412 | 0.4124 | 0.4127 |
|  | Wave | 0.4087 | 0.412 | 0.4124 | 0.4127 |

We received the same results for message saturation time when the simulations were run with 5000KB node buffers. This again makes intuitive sense: as no messages were removed when buffers were 500KB in size, messages also are not removed when buffers are 5000KB in size. Average node buffer capacity results were nearly identical to the results of the 500KB buffer simulations, as one would expect, just out of a total capacity of 5000KB instead of 500KB. We expect the ONE simulator report used to measure buffers rounds or truncates to four decimal places, and this is the cause of the small discrepancies. These results are shown in Table 3.

TABLE III.  MAX AVERAGE BUFFER OCCUPANCY BY NODE COUNT AND ROUTING PROTOCOL WITH NODES WITH 5000KB BUFFERS.

|  |  | 100 Nodes | 500 Nodes | 1000 Nodes | 3000 Nodes |
|---|---|---|---|---|---|
| Max Average Buffer Occupancy (% of 5000KB) | Epidemic | 0.0409 | 0.0412 | 0.0412 | 0.0413 |
|  | Wave | 0.0409 | 0.0412 | 0.0412 | 0.0413 |

### B. Moderate Message Frequency Simulations

1 message per NOI created at the beginning of the simulation; then, 1 message created per NOI at least every 5 minutes. In all simulations, all messages again reached 100% saturation. Figure 9 shows the rolling average of the time it takes for a message to reach 100% saturation by the time the message was created for Epidemic and 500KB node buffers. A β of 0.9 is used (represented as α = 0.1 in the calculation code). Figure 10 shows the same for the Wave protocol.

Epidemic (500KB buffers)

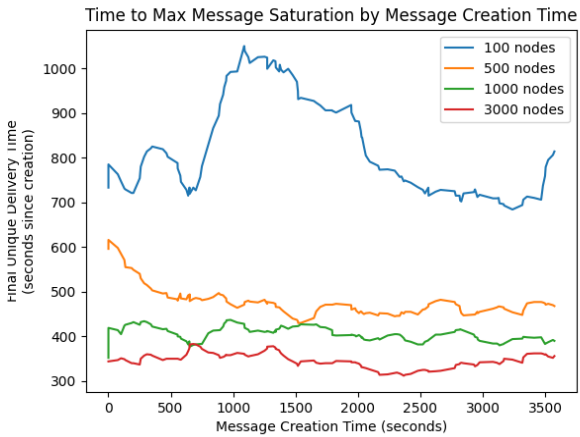

Fig. 9. Moving average of the time elapsed between message creation and the time at which a message has saturated all nodes. Epidemic protocol with 500KB node buffers. β = 0.9 (α = 0.1).

Wave (500KB buffers)

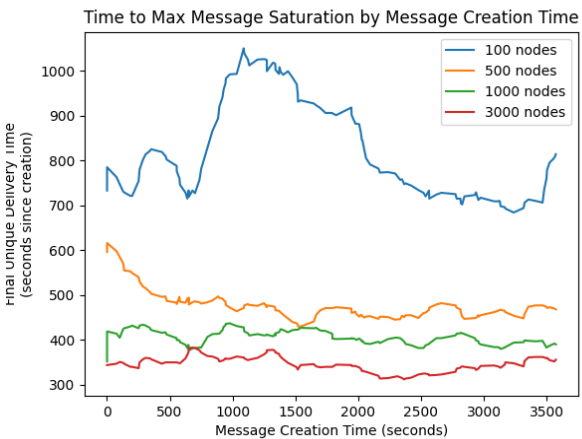

Fig. 10. Moving average of the time elapsed between message creation and the time at which a message has saturated all nodes. Wave protocol with 500KB node buffers. β = 0.9 (α = 0.1).

We note that when viewed in a scatterplot, there is a significant outlier in the 100 node simulations around the 1000s mark that notably shifts the moving average. We leave normalizing data for future works.

While notably more messages were created than in the prior simulations, node buffers still did not come close to reaching capacity, with the 500KB buffers reaching a maximum average occupancy of ~50% and the 5000KB buffers expectedly ~5%. These results are shown in Table 4 and Table 5.

TABLE IV.  MAX AVERAGE BUFFER OCCUPANCY BY NODE COUNT AND ROUTING PROTOCOL WITH NODES WITH 500KB BUFFERS.

|  |  | 100 Nodes | 500 Nodes | 1000 Nodes | 3000 Nodes |
|---|---|---|---|---|---|
| Max Average Buffer Occupancy (% of 500KB) | Epidemic | 47.7775 | 49.9703 | 50.5201 | 50.9925 |
|  | Wave | 47.7775 | 49.9703 | 50.5201 | 50.9925 |

TABLE V.  TABLE 5: MAX AVERAGE BUFFER OCCUPANCY BY NODE COUNT AND ROUTING PROTOCOL WITH NODES WITH 5000KB BUFFERS.

|  |  | 100 Nodes | 500 Nodes | 1000 Nodes | 3000 Nodes |
|---|---|---|---|---|---|
| Max Average Buffer Occupancy | Epidemic | 4.7777 | 4.997 | 5.052 | 5.0992 |
|  | Wave | 4.7777 | 4.997 | 5.052 | 5.0992 |

We include graphs of the moving averages of time to max message saturation for 5000KB node buffers in Figure 11 and Figure 12. As neither the 500KB nor the 5000KB buffers reach capacity, it makes sense we again see shared behavior in the shape of message saturation graphs and buffer use across both protocols and buffer sizes.

Epidemic (5000KB buffers)

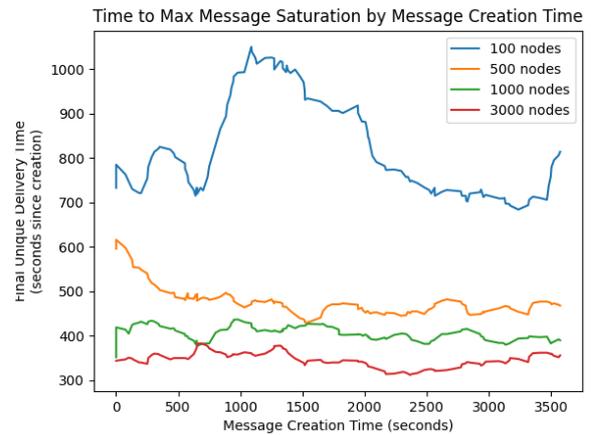

Fig. 11. Moving average of the time elapsed between message creation and the time at which a message has saturated all nodes. Wave protocol with 5000KB node buffers. β = 0.9 (α = 0.1).

Wave (5000KB buffers)

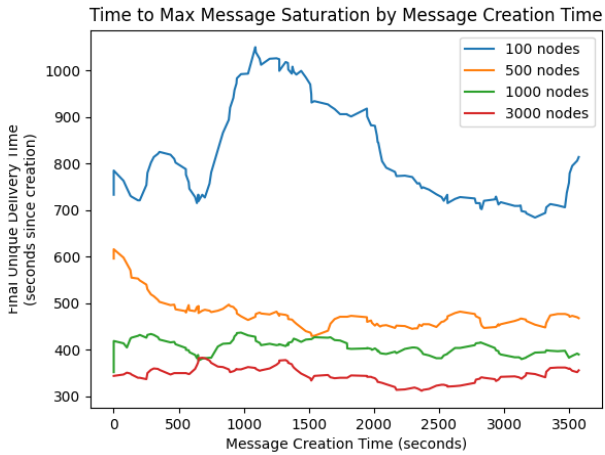

Fig. 12. Moving average of the time elapsed between message creation and the time at which a message has saturated all nodes. Wave protocol with 5000KB node buffers. β = 0.9 (α = 0.1).

*C. High Message Frequency Simulations*

1 message per NOI is created at the beginning of the simulation; then, 1 message is created per NOI at least every 30 seconds. Simulations with messages created at this frequency and high node counts, especially the 3000 node simulations, were computationally taxing. We ran our simulations on a 2021 M1 Pro MacBook Pro with 16GB of RAM, and this proved to be insufficient compute for some simulations. For example, after running an Epidemic simulation with 3000 nodes for ~14 hours, only 2425 seconds of the 9000 second simulation had processed. In another instance, we encountered a Java OutOfMemory error after 1676 seconds of the simulation had processed. Because of this, there are some results we were unable to gather completely. We identify such results with a hyphen where relevant. Additionally, we recommend future works exploring networks of this scale procure more compute to efficiently run the simulations.

Figure 13 and Figure 14 show the max average buffer capacities of nodes with 500KB buffers. We were unable to fully gather data for the Epidemic 3000 node simulation due to compute constraints.

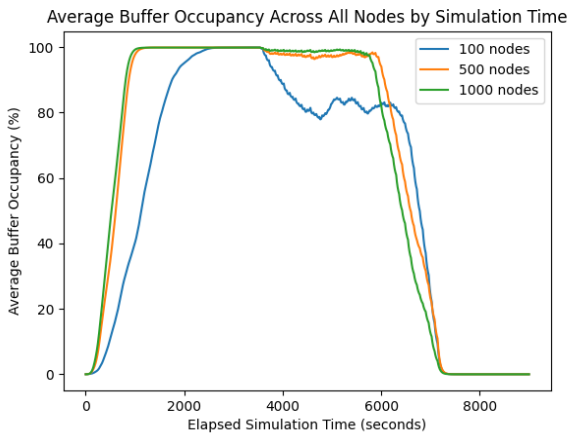

Fig. 13. Average buffer occupancy across all nodes by simulation time. Epidemic routing with 500KB node buffers.

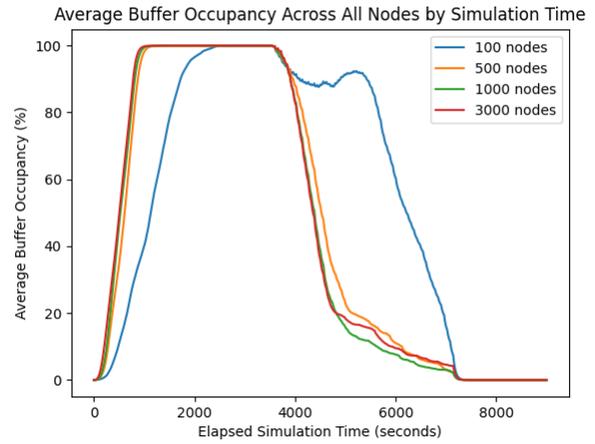

Fig. 14. Average buffer occupancy across all nodes by simulation time. Wave routing with 500KB node buffers.

Unlike previous simulations, node buffer occupancy averaged near maximum capacity in all simulations for notable amounts of simulation time. While both routing protocols saw node buffers eventually decrease from near maximum occupancy, the Wave protocol saw this decrease earlier than the Epidemic protocol for the 500 and 1000 node simulations. The Wave protocol simulations saw buffer occupancy begin to notably decrease between 4000 and 6000 seconds of simulation time, while the corresponding Epidemic protocol simulations saw the notable decrease begin between 6000 and 8000 simulation seconds.

Maximum message saturations differed from previous simulations as well. Figure 15 and Figure 16 present scatterplots of maximum saturation percent of messages by message creation time.

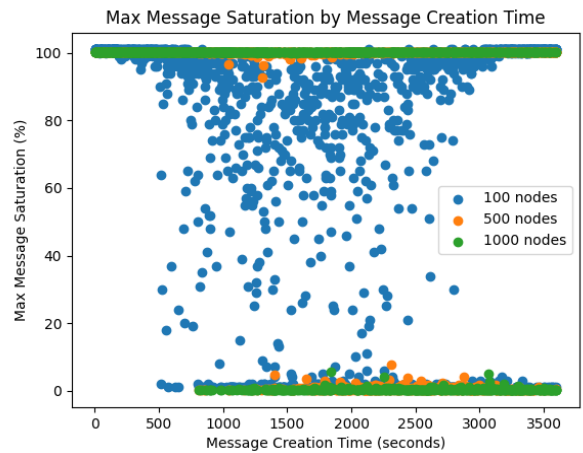

Fig. 15. Maximum saturation percentage of a message by the time the message was created. Epidemic routing with 500KB node buffers.

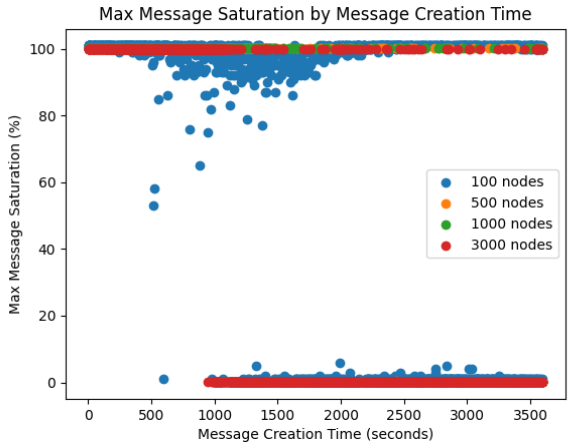

Fig. 16. Maximum saturation percentage of a message by the time the message was created. Wave routing with 500KB node buffers.

While messages in the 500, 1000, and 3000 node simulations generally achieved either near 100% saturation or near 0% saturation, messages in the 100 node simulations experienced a broader range of saturation values. Counts of messages which did not reach 100% saturation by simulation are presented in Table 6. As data for the 3000 node Epidemic simulation was unable to be gathered, we have omitted it.

TABLE VI. COUNT OF MESSAGES WHICH DID NOT REACH 100% SATURATION BY ROUTING PROTOCOL AND QUANTITY OF NODES. 500KB NODE BUFFERS.

| | | 100 Nodes | 500 Nodes | 1000 Nodes | 3000 Nodes |
|---|---|---|---|---|---|
| Count of Messages <100% Saturation | Epidemic | 884 | 391 | 382 | - |
| | Wave | 615 | 745 | 801 | 814 |

In the 100 node simulations, the Epidemic protocol saw 884 messages unable to reach 100% saturation while Wave saw only 615. However, in the 500 and 1000 node scenarios, Epidemic saw notably fewer messages unable to reach 100% saturation than Wave. While Wave saw 745 and 801 messages unable to reach 100% saturation, respectively, Epidemic saw fewer than 400 messages unable to reach 100% saturation in both simulations. Interestingly, as the number of nodes increased, the Wave protocol saw more messages unable to reach 100% saturation while the Epidemic protocol saw fewer.

Of the messages that did reach 100% saturation, Figure 17 and Figure 18 show the time it took a message to reach 100% saturation by the time the message was created.

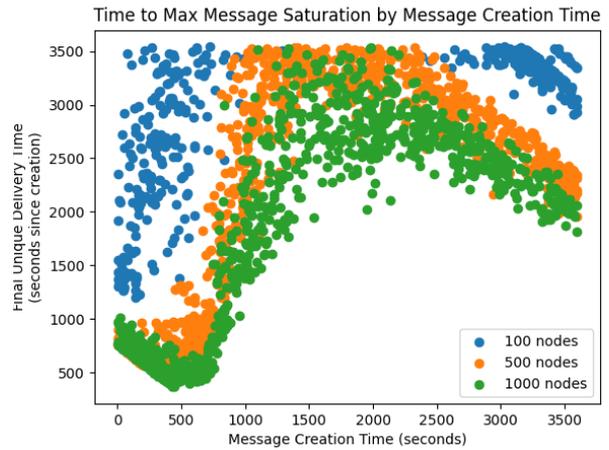

Fig. 17. Time it takes a message to achieve 100% saturation by message creation time. Only includes messages which reached 100% saturation. Epidemic routing with 500KB node buffers.

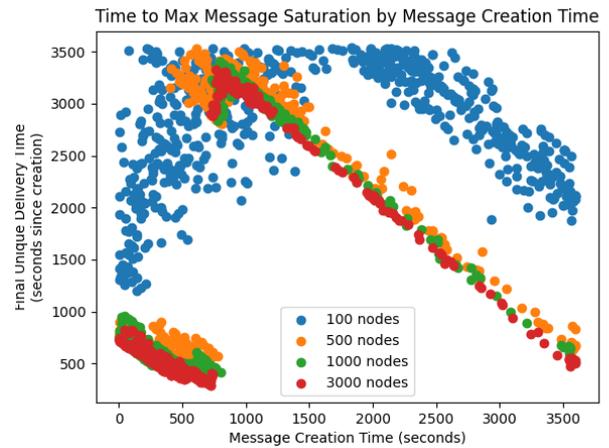

Fig. 18. Time it takes a message to achieve 100% saturation by message creation time. Only includes messages which reached 100% saturation. Wave routing with 500KB node buffers.

Both protocols saw a sharp rise in the amount of time it takes a message to saturate the network early in the simulation and then a decrease afterwards. However, the rates at which the decrease occurs differs, with the Wave protocol showing a sharper decrease than Epidemic.

In the simulations with 5000KB node buffers, the buffers did not average near maximum occupancy in any simulation we could process. We present their max average buffer occupancies in Table 7.

TABLE VII. MAX AVERAGE BUFFER CAPACITY BY NODE COUNT AND ROUTING PROTOCOL WITH NODES WITH 5000KB BUFFERS.

| | | 100 Nodes | 500 Nodes | 1000 Nodes | 3000 Nodes |
|---|---|---|---|---|---|
| Max Average Buffer Capacity (% of 5000KB) | Epidemic | 26.9496 | 49.3453 | 49.7643 | - |
| | Wave | 26.9496 | 49.3453 | 49.7643 | - |

As in prior simulations in which buffers did not reach average occupancies near their maximum capacity, we see shared results between Epidemic routing and Wave routing. However, a surprising difference from prior simulations is that, despite node buffers not averaging near maximum capacity, many messages did not reach 100% saturation in either of the 100 node simulations. This is shown in Figure 19 and Figure 20.

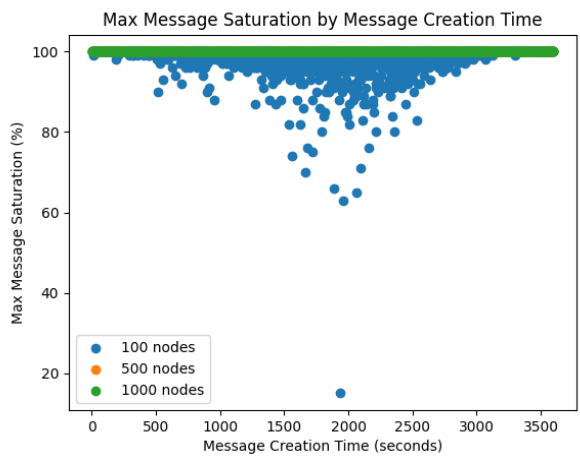

Fig. 19. Maximum message saturation (number of nodes delivered to / total number of nodes in network) by message creation time. Epidemic routing with 5000KB node buffers.

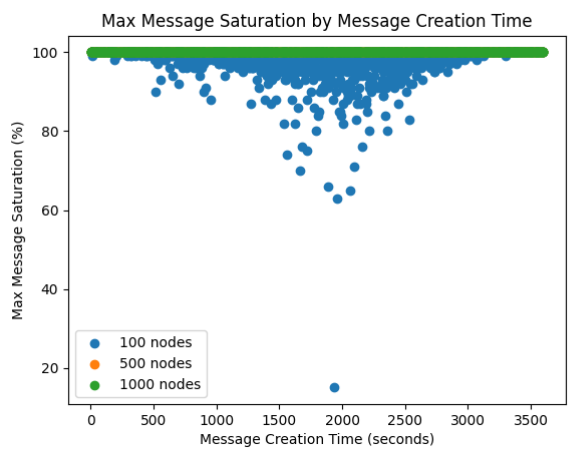

Fig. 20. Maximum message saturation (number of nodes delivered to / total number of nodes in network) by message creation time. Wave routing with 5000KB node buffers.

Specifically, 659 messages did not reach 100% saturation in both 100 node simulations. While this is unexpected, we do also observe that the result is still consistent between the protocols. We recommend further works explore this finding further.

## V. Discussion

For our scenario configuration, the differences in behavior between networks which implement the Epidemic routing protocol and networks which implement the Wave routing protocol begin to show when node buffers begin reaching maximum occupancy.

In simulations in which only one message is generated, we see shared results between Epidemic routing and Wave routing for simulations of all node quantities: the saturation percentage by elapsed simulation time graphs are of the same shape, and the times it takes for the message to reach 100% saturation are also the same.

This makes intuitive sense. In our implementation, the primary differentiator between the protocols is the Wave protocol's immunity time. When a message is removed from a node's buffer in a Wave network, the immunity time will prevent the message from being added back before its TTL expires; for Epidemic, this is not the case. Accordingly, when node buffers do not reach maximum occupancy and no messages are removed other than for TTL expiration, it is understandable that we do not see a difference in the behavior of the protocols.

This identification is reinforced by the results we see from the Moderate Message Frequency simulations and the High Message Frequency simulations with 5000Kb node buffers: node buffers once again do not reach an average occupancy near maximum capacity, and once again we see shared shapes between message saturation graphs.

In the High Message Frequency simulations with 500KB node buffers, we see differences between the behaviors of networks which implement the different protocols. After node buffers' average occupancy reaches near maximum capacity, buffer occupancy percentage decreases notably more quickly in the Wave simulations than the Epidemic simulations. We expect this is a result of the Wave protocol's immunity time: in our Wave networks, when a message is removed from a node's buffer because the buffer is at capacity, the node will not accept that message again later on when buffer space is available again. In networks which implement Epidemic routing, the node would accept the message again in this type of situation.

Practically, this finding suggests it could be beneficial to use our scenario's implementation of the Wave routing protocol in networks where node buffers are expected to reach maximum occupancy and message re-delivery is not necessary. In relation to our MacArthur Park scenario specifically, a messaging app could store a copy of a message in persistent storage outside of the node buffers, in which case the faster decrease in node buffer occupancy exhibited here could be highly desirable.

Another difference between the protocol behaviors, and one we find surprising, is in the counts of messages which do not reach 100% message saturation in the High Message Frequency simulations. We expected Wave protocol networks to have less competition for node buffer space than Epidemic protocol networks due to the Wave protocol's immunity time, and accordingly we hypothesized that fewer messages would fail to reach 100% saturation in Wave networks than Epidemic networks. However, in two of the three simulations for which we can compare data, the data shows the opposite: significantly fewer messages fail to reach 100% saturation in the Epidemic networks than the Wave networks. In other words, in two of the three simulations, more messages are delivered to every node in the network when the network uses Epidemic routing instead of Wave routing.

However, when evaluating only messages that do reach 100% saturation in these simulations, there is a difference between the protocols in the time it takes a message to reach that point. Both protocols experience a sharp increase in the time it takes a message to reach 100% saturation early in the simulation; then, the Wave protocol sees a sharper decrease in time to 100% saturation than Epidemic. In other words, after the sharp increase in time to 100% message saturation, messages in the Wave networks begin saturating the network more quickly again than in the Epidemic networks. We expect this also results from our Wave protocol implementation's immunity time: with removed messages unable to re-occupy buffer space, more buffer space is available for new messages.

Practically, these findings suggest that if a message saturating the network is the highest priority, it can be more beneficial to use Epidemic routing. Conversely, in a situation where it is not important for every message to saturate a network but those that do should do so quickly, it can be more beneficial to use Wave routing. Regarding our MacArthur Park scenario, these findings indicate it may be more beneficial to use Epidemic routing in a highly-populated situation than Wave routing if the top priority is that others nearby receive information about the acts of questionable legality occurring. We recommend further works explore these phenomena further and at a more granular level, as we find the first one described especially surprising.

We also want to discuss two additional phenomena we observed. First, in the Event Log Reports generated by ONE for the Wave protocol simulations, we observed entries indicating that messages were re-delivered to nodes. As discussed, we do not believe this should be possible with our configuration of the Wave protocol's immunity time. We recognize this phenomena does not necessarily mean the message is being treated as redelivered in practice, as this could be resulting from something such as a discrepancy between how the Event Log Report records data and how ONE's WaveRouter Java class operates. We recommend future works explore this phenomena and the ONE's source code in detail to identify the cause.

Second, when evaluating max message saturation percentage, we sometimes saw saturation percentages greater than 100% appear, such as 100.3%. We suspect this due to a situation similar to the following occurring:

1) A node creates a new message.

2) The created message is eventually removed from the node's buffer to make room for another new or incoming message.

3) The node eventually has available buffer space and the message created in 1) is re-delivered to the node.

This is another phenomenon we would not have expected to be possible with our set immunity time. We recommend future works explore this further, including within the ONE simulator's source code.

## VI. ADDITIONAL RECOMMENDATIONS FOR FUTURE RESEARCH

In addition to the recommendations for future research we have made throughout this work, we also recommend that future works explore the behavior of BitChat's implementation of "Efficient Gossip with Bloom Filters" [35] in our MacArthur Park scenario. This will enable the simulation of an even more realistic scenario and may provide valuable data for users, developers, and other researchers regarding effective means of communication in such a situation.

We also recommend exploring communication at protests, which are another situation where efficient, secure communication is of high importance as discussed by [2], [20]. In addition to the user-to-user / group messaging largely discussed in [2], [20], we also see a need for efficient distribution of a message to all nodes in the network. For example, if chemical agents such as tear gas are used against protestors, it may be important for protestors not in the area of deployment to be aware of its use to avoid the area. As [2], [20] discuss Epidemic-like protocols, such a goal may be able to be evaluated with few or no modifications to the protocols discussed in those works.

New York City (New York, New York) has had multiple large protests in 2025 and we believe could be an appropriate location for evaluation. The New York Police Department stated that more than 100,000 people protested across the city's five boroughs at the No Kings protest on October 18th, 2025 [56]. This makes the city, and more specifically the location of the October 18th, 2025 No Kings protest march, a setting for analyzing message saturation rates in large, dense protest settings.

Lastly, we recommend the exploration of a, we believe novel, Epidemic-like protocol which considers a node's distance from other nodes when routing a message. When selecting which other node to replicate a message to, a given node will alternate between replicating the message to the node closest to it which does not already have the message and the node furthest from it which does not already have the message. We hypothesize this may result in a message being delivered to every node in the network more quickly than other protocols, and may also help reduce any local congestion that occurs when messages are distributed in more naïve manners. We leave our exploration of such a protocol for a future work.

## VII. OTHER WORKS

MANETs and DTNs are also being used and explored for many purposes outside of those explored and previously discussed in this work. DTNs can be used in Vehicular Ad Hoc Networks (VANETs), special types of MANETs specifically for communication amongst vehicles and infrastructure [57], to transmit safety information and other types of data [58].VANETs and DTNs can also be used to provide internet access to locations where accessing the internet by traditional means may be unavailable, such as rural areas [58]. DTNs can be used for low earth orbit and deep space communications [12], [59]. There are many works which investigate different aspects of this use case. DTNs can be used for agricultural and public health purposes, such as for livestock monitoring [60]. These are just a few of the many further applications of MANETs and DTNs. We leave further exploration to the reader.

## VIII. CONCLUSION

In this work, we explored the behavior of the Epidemic and Wave DTN routing protocols in a realistic setting where individuals may wish to communicate with others for support regarding an act of questionable legality. We identified situations where using the Epidemic routing protocol may be more advantageous in such a scenario, and situations where using the Wave routing protocol may be more advantageous instead. We also discussed other aspects of our findings in detail and suggested multiple approaches to future works.